\documentclass[11pt]{article}
\usepackage{geometry}                
\usepackage{graphicx}
\usepackage{amssymb}
\usepackage{amsmath}
\usepackage{mathtools}
\usepackage{epstopdf}
\usepackage{bm}
\usepackage{bbm}
\usepackage[auth-sc]{authblk}
\newcommand{\be}{\begin{eqnarray}}
\newcommand{\ee}{\end{eqnarray}}
\newcommand{\la}{\langle}
\newcommand{\ra}{\rangle}
\newcommand{\tr}{\rm Tr}

\title{Emergence of functional information from multivariate correlations}

\author[1,2,3,4,*]{Christoph Adami}
\author[2,5]{Nitash C G}
\affil[1]{Department of Microbiology and Molecular Genetics}
\affil[2]{BEACON Center for the Study of Evolution Action}
\affil[3]{Department of Physics and Astronomy}
\affil[4]{Program in Ecology, Evolution, and Behavior}
\affil[5]{Department of Computer Science and Engineering}
\affil[ ]{ Michigan State University, East Lansing, MI 48824}
\affil[*]{adami@msu.edu}

\date{}   
\begin{document}                                        
\maketitle
\begin{abstract}
The information content of symbolic sequences (such as nucleic- or amino acid sequences, but also neuronal firings or strings of letters) can be calculated from an ensemble of such sequences, but 
because information cannot be assigned to single sequences, we cannot correlate information to other 
observables attached to the sequence. Here we show that an information {\em score} obtained from multivariate (multiple-variable) correlations within sequences of a ``training" ensemble can be used to predict observables of out-of-sample sequences with an accuracy that scales with the complexity of correlations, showing that functional information emerges from a hierarchy of multi-variable correlations.  
\end{abstract}

\section{Introduction}
The term ``functional information" refers to the information that is necessary and sufficient to predict the function of a symbolic sequence within a particular environment. Typically, the sequence could be nucleic or amino acids, but it also could be a sequence of neuronal firings or even a string of letters written in a particular alphabet. The function could be the biochemical activity (for biological sequences), or the level of performance for a neuronal network that produces the firings. For words in a language, the functionality could be simply the likelihood that the word is understood by someone familiar with the language. In some sense, we often simply use the word ``information" to indicate functional information. For example, the hemoglobin DNA sequence carries the information necessary to make the protein hemoglobin  (given the right cellular environment) is functional in the sense that it binds oxygen with a particular affinity. For recordings from animal brains, the functionality of the sequence of neuronal activations could be the accuracy with which the animal performs a rewarded task. Functional information can be defined quantitatively in terms of the likelihood that a {\em random} sequence performs the function or task at the specified level~\cite{Szostak2003,Hazenetal2007}. Specifically, if we require a particular sequence to   perform function $E$ at least at the level $\theta$, then functional information about $E$ as defined by Szostak is given by
\be
I(E_\theta)=-\log F(E\geq\theta)\;, \label{fun}
\ee
where $F(E\geq\theta)$ is the fraction of {\em all} sequences (say, given a particular length $n$) that perform the function at a level at least $\theta$.

Such a definition of functional information makes eminent sense (it implies that functional sequences are exponentially rare) and accurately links information content to functionality. For example, in the evolution of RNA molecules that encode an enzymatic activity~\cite{Carothersetal2004}, functional information is correlated to both activity and structural complexity.  In the evolution of HIV protease~\cite{GuptaAdami2016} functional information can be seen to decrease when anti-viral drugs are given (reflecting the reduced activity of the protease), then to increase again as anti-viral resistance emerges. In fact, functional information defined in this manner turns out to be just the coarse-grained Shannon information content of the sequences~\cite{AdamiLaBar2017}.  

A shortcoming of this measure is that only large groups of sequences can be characterized in this manner, because accurately estimating the fraction $F(E\geq\theta)$ necessitates large  ensembles of functional sequences. Here we develop an approach that allows us to attach a score to {\em every individual sequence} in an ensemble, which then allows us to predict the level of functionality of sequences {\em not} in the ensemble. In particular, this approach allows us to study how functional information emerges from multi-site correlations within the sequence.

\section{Information content of functional symbolic sequences}
We now define the information content of a set of sequences $\vec s^{(k)}$ of length $L$ that form an ensemble ${\cal S}$, given a particular environment within which they aqre functional~\cite{AdamiCerf2000,Adami2004,Adami2012}. The sequences $\vec s^{(k)}$ are written in terms of the state of a joint random variable $S=S_1S_2\cdots S_L$ (one for each monomer), that is, $\vec s^{(k)}=(s_1^{(k)},s_2^{(k)},\cdots,s_L^{(k)})$, while the environment can be thought of as the state of an environment variable, that is, $E=e$. With this notation, the information content 
can be written as
\be
I(S:E=e)=H(S)-H(S|e)\;. \label{info}
\ee
In Eq.~(\ref{info}), we defined two ensembles: the unconditional (unconstrained) ensemble ${\cal S}$ and the constrained (by function within environment $e$) ensemble ${\cal S}_e$. Typically, the unconstrained ensemble is all the possible sequences of a given length $n$, while the ensemble ${\cal S}_e$ consists of only those sequences that are functional in $e$ at a specified level. Further, $H(S)$ is the Shannon entropy of the unconstrained ensemble which, if sequences of length $L$ are drawn from an alphabet of $D$ symbols, is simply $H(S)=L\log D$. The entropy of the constrained ensemble requires the knowledge of the likelihood $p_k$ to find a functional sequence $\vec s^{(k)}$ within the ensemble. Then, we can calculate the Shannon entropy of sequences in ${\cal S}_e$ as
\be
H(S|e)=-\sum_k^{N_e}p_k\log p_k\;, \label{ent}
\ee
where $N_e$ is the number of different functional sequences within ${\cal S}_e$. 
If we coarse-grain entropy (\ref{ent}) by assuming that each sequence appears with equal probability, then $p_k=1/N_e$ and
\be
I(S:e)=\log D^L-\log N_e=-\log\left(\frac{N_e}{D^L}\right)\;,
\ee
which is Szostak's functional information since $F=\frac{N_e}{D^L}$. 

Because most functional sequences of interest are exponentially rare, finding a sufficient number of them to accurately estimate the fraction $F(E\geq\theta)=N_e/D^L$ is difficult or nearly impossible. For example, while tens of thousands of variants of the 99-mer HIV-1 protease are known, they are not obtained by randomly sampling the 99-mer space, and therefore cannot be used to calculate functional information in this manner. However, it is possible to relate the information $I(S:e)$ to monomer, dimer, and tri-mer entropies (and so on) as follows.

In general, a multi-variable entropy such as $H(S|e)=H(S_1S_2\cdots S_L)$ can be decomposed into multi-mer correlation entropies using Fano's entropy and information decomposition theorems~\cite{Fano1961}). Specifically\footnote{Fano's decomposition theorems are just specific information-theoretic examples of the more general inclusion-exclusion theorem that has been known for some time. For example~\cite{Sylvester1883} refers to it as a ``th\'eor\`eme logique bien connu".}
\be
H(S_1\cdots S_L)=\sum_{m=1}^L(-1)^{m-1}\sum_{\sigma(1)\cdots\sigma(m)}^nI(S_{\sigma(1)}:\cdots: S_{\sigma(m)})\;. \label{entdecomp}
\ee
and 
\be
I(S_1:\cdots: S_L)=\sum_{m=1}^L(-1)^{m-1}\sum_{\sigma(1)\cdots\sigma(m)}^LH(S_{\sigma(1)}\cdots S_{\sigma(m)})\;. \label{infodecomp}
\ee
Here, we write $I$ for informations (shared entropies) for ease of reading. In these formul\ae, $\sigma(i)$ stands for a permutation of the index $i$. 
For example, the entropy decomposition theorem (\ref{entdecomp})
can also be written as
\be
H(S_1\cdots S_L)=\sum_{i=1}^L H(S_i)-\sum_{i<j}^LI(S_i:S_j)+\sum_{i<j<k}^LI(S_i:S_j:S_k)-\cdots(-1)^{L-1}I(S_1:\cdots: S_L)\;. \label{entdecomp2}
\ee
These theorems allow us to approximate the exact information content $I(S:e)$. To start with, if we neglect all correlations between variables, then the first-order approximation to $H(S|e)$ is
\be 
H_1(S|e)=\sum_{i=1}^LH(S_i)\;, \label{ent1}
\ee
where $H(S_i)$ is the (functional) monomer entropy of site $i$, that is,
\be
H(S_i)=-\sum_{s_i}p_i(s_i)\log p_i(s_i)\;, \label{single-ent}
\ee
with $p_i(s_i)$ the likelihood to find symbol $s_i$ (one out of $D$ possible ones) at position $i$.

If we keep correlations between pairs of sites but neglect correlations between three sites, we obtain the second-order approximation of the entropy as
\be
H_2(S|e)=\sum_{i=1}^LH(S_i)-\sum_{i<j}^L I(S_i:S_j)\;, \label{ent2}
\ee
where $I(S_i:S_j)$  is the entropy shared between variables $S_i$ and $S_j$. We can deduce how this is written in terms of marginal entropies (entropies obtained by summing over one or more of the monomers) using the decomposition theorem (\ref{infodecomp}) for $L=2$ as
\be
I(S_i:S_j)=H(S_i)+H(S_j)-H(S_iS_j)\;,
\ee
the standard Shannon formula for information. 

Because $H(S_i)$ and $I(S_i:S_j)$ can be estimated from finite ensembles of functional sequences, approximations such as (\ref{ent2}) are sufficient to estimate the information content $I(S:e)$ as long as correlations between three or more variables can be neglected (see, e.g.,~\cite{GuptaAdami2016}).

While these approximations are useful for calculating the information content of sequences from finite ensembles, they do not allow us to correlate individual characters of a sequence (for example, its particular level of functionality) to information because, as we have emphasized repeatedly, information is not a property of a single sequence. However, each sequence is also a {\em pattern}, so it should be possible in principle to relate the individual pattern in the sequence to its function. Indeed, methods of this sort have been developed before: they use sophisticated techniques to link
the patterns to function, either by regression~\cite{Hinkleyetal2011,Kouyosetal2011} or by using a multiple-sequence alignment (MSA) to train a probabilistic model (usually called ``Potts" or ``Ising" model)~ \cite{Schneidmanetal2006,MoraBialek2011,Fergusonetal2013,Biswasetal2021}. We now show that there is a much simpler approach that allows us to link patterns to function directly, without regression or fitting a Potts model. 

\section{Model-free linkage of pattern to function} 
A common technique to model correlations between variables in a sequence $\vec s=(s_1,s_2\cdots,s_L)$ is the so-called Potts or Ising model Hamiltonian (energy functional)\footnote{Techically speaking, the model is neither a Potts nor an Ising model, as those models only have nearest-neighbor interactions. It is best called an infinite-range Ising-like model.} that can be written as~\cite{Schneidmanetal2006,MoraBialek2011}
\be
E(\vec s)=\sum_{i=1}^L h_i(s_i)+\sum_{i<j}^LJ_{ij}(s_i,s_j)\;. \label{ham1}
\ee
Here as before, the $s_i$ are the sequence monomers taken from an alphabet of size $D$. For neuronal firings, the $s_i$ are binary, while for much of the literature on molecules, the $s_i$ are reduced to a small alphabet, for example $s_i=0$ for wild-type monomer, and $s_i=1$ for a mutant.  The  $L\cdot D$ parameters $h_i(s_i)$ and ${L\choose 2}D^2$ parameters $J_{ij}(s_i,s_j)$ are Lagrange multipliers obtained by maximizing the entropy 
\be
H(\vec s)=-\sum_{k=1}^{D^L}p_k\log p_k\;, \label{Pottsent}
\ee
constrained to reproduce the monomer and dimer probabilities $p(s_i)$ and $p(s_i,s_j)$ that can be obtained from the MSA. The probability $p_k=P(\vec s^{(k)})$ is incidentally the same as defined in (\ref{ent}): it is the likelihood to find sequence $\vec s^{(k)}$ by chance. Maximization under the constraints yields the familiar Boltzmann distribution
\be
p_k=\frac 1Ze^{-E(\vec s)}\;, \label{bol}
\ee
with $E(\vec s)$ from (\ref{ham1}) above, and where
\be
Z=\sum_{k}e^{-E(\vec s^{(k)})}
\ee
is the partition function.

Now, let's take a closer look at the entropy $H(\vec s)$. Using (\ref{bol}), we can write this as
\be
H(\vec s)=\log Z+\sum_{i=1}^L\sum_{s_i}h(\sigma_i)p(s_i)+\sum_{i<j}\sum_{s_i,s_j}p(s_i,s_j)J_{ij}(s_i,s_j)\;. \label{entpotts}
\ee
It is clear that the right-hand-side of Eq.~(\ref{entpotts}) is an approximation of the ``true" entropy $H(\vec s)$, fitted to the MSA using the first and second-order Lagrange multipliers. Higher-order terms are neglected. But it is a strange formula, because $\log Z$ also appears on the right-hand-side, and we don't have an approximation for that. But it turns out that we can approximate the left-hand-side using averages taken from the MSA as well, up to any order we want. This is possible because we can decompose $H(\vec z)$ exactly using the entropy decomposition theorem, for example written in the form~(\ref{entdecomp2}). Using this approximation, we will be able to determine $\log Z$, and eliminate the Lagrange multipliers.

Because the Potts model parameters only consider up to bi-partite correlations, we will neglect terms of order three or higher in (\ref{entdecomp2}) for now, but it is clear that we could keep them as long as we fit an appropriate third-order term in the Hamiltonian (\ref{ham1}). Using $p_i(s_i)$ directly from the MSA, along with the $p_{ij}(s_i,s_j)$, and so forth, we can 
insert those terms in (\ref{entpotts}) to yield
\be
\log Z&=&-\sum_i\sum_{s_i}p_i(s_i)\biggl(\log p_i(s_i)+h(s_i)\biggr)
+\sum_{i<j}\sum_{s_i,s_j}p_{ij}(s_i,s_j)\biggl(\log  \frac{p_i(s_i)p_j(s_j)}{p_{ij}(s_i,s_j)}-J_{ij}\biggr)\nonumber\\
&=&\sum_i\biggl(H(i)-\la h_i\ra\biggr)-\sum_{i<j}\biggl(H(i:j)+\la J_{ij}\ra\biggr) \label{part}
\ee
 In (\ref{part}), we defined the average Lagrange coefficients
 \be
\la h_i\ra=\sum_{s_i}p_i(s_i)h(s_i)
\ee
and 
\be
\la J_{ij}\ra=\sum_{s_i,s_j}p_{ij}(s_i,s_j)J_{ij}(s_i,s_j)\;.
\ee
We also introduced the simplified notation $H(i)\equiv H(S_i)$ and $H(i:j)\equiv H(S_i:S_j)$ and so on, which we will use from now on.

In order to understand the term $\log Z$, we turn to Berg-von Hippel (BvH) theory~\cite{BergvonHippel1987,BergvonHippel1988}. BvH theory defines an energy functional  for a DNA sequence that is related to the binding affinity of a transcription factor using a MSA. Indeed, BvH theory is very similar to the max-entropy approach from~\cite{MoraBialek2011}, except that it assumes that the distribution {\em at each site} is at maximum entropy.  In fact, the energy score of a DNA binding site is given precisely by an energy functional of the form (\ref{ham1}), but without the second-order term (as correlations between sites are close to irrelevant in DNA binding). 

Unlike in the Potts-model approach, the counterparts to $h_i(s_i)$ are not fitted to the MSA in the BvH approach. Rather, they are directly extracted from the MSA in terms of a matrix as\footnote{We changed notation slightly in Eq.~(\ref{pwm}), from $p_i(s_i)$ in Eq.~(\ref{single-ent}) to $p_i(s_a)$ here, to conform to BvH theory. The difference is that in BvH theory it is convenient to index the symbol, so that it is obvious that $p_i(s_a)$ is a matrix element, while this was implicit in the form $p_i(s_i)$. In the latter formulation, rather than summing over an index we sum over the symbols $s_i$.}
\be
M^{(1)}_{ia}=\lambda \log\frac{p_i(s_0)}{p_i(s_a)}\;, \label{pwm}
\ee
because this is the expression that maximizes the single-site entropy (subject to constraints). We can see this by writing 
\be
p_i(s_a)=p_i(s_0)e^{-\frac{M^{(1)}_{ia}}{\lambda}}\;.
\ee
In Eq.~(\ref{pwm}), $\lambda$ is a Lagrange multiplier that allows us to set the scale of the energy just as we did in the earlier maximization (we set $\lambda=1$ here). Furthermore,
$p_i(s_0)$ is the probability (likelihood) to find the consensus monomer at site $i$. Obviously, the consensus monomer is the one that has the highest frequency in the MSA at that position. This assures that the consensus sequence is the one with vanishing energy (the ``ground state").  The matrix $M^{(1)}$ is called the ``position-weight matrix" (PWM) of a sequence~\cite{Stormo2000}, and it is fairly successful at predicting DNA binding sites in a genome (see, e.g.,~\cite{BrownCallan2004}), and even at detecting correlations between binding sites~\cite{CliffordAdami2015}.

We now introduce a second-order PWM (as BvH in fact did in the appendix to~\cite{BergvonHippel1987}, but for adjacent monomers only)
\be
M^{(2)}_{ijab}=\log\biggl(\frac{p_{ij}(s_0)}{p_{ij}(s_a,s_b)}\biggr)\;. \label{pwm2}
\ee
Just as before, $p_{ij}(s_a,s_b)$ is the likelihood to find symbol combination $s_a,s_b$ at the pair of sites $i,j$, while $p_{ij}(s_0)$ is the likelihood to find the consensus pair there\footnote{A more precise (but also more cumbersome) notation would be $p_{ij}(s_0^{(ij)})$, to indicate that in pair-wise probabilities, $s_0$ refers to the consensus {\em dimer}.}. 
Defining this matrix implicitly assumes that any pair of monomers is also in statistical equilibrium. While assuming that even higher-order correlations are in statistical equilibrium might be increasingly doubtful, it is certainly more likely than assuming that the entire sequence is at maximum entropy, as (\ref{bol}) does.

We can now use BvH theory to calculate $\log Z$ and insert it into Eq.~(\ref{part}). In standard BvH theory, $\log Z$ is given by (recall we are setting $\lambda=1$ here)
\be
\log Z=-\sum_{i=1}^L \log p_i(s_0)\;,
\ee
but this is only the first-order approximation. In general, for any sequence $\vec s$ we must have
\be
Z=\frac1{p_0(s_1\cdots s_L)}\;,
\ee
which we can expand so that\footnote{This is the expansion of a joint probability that is at the root of Fano's entropy and information decomposition theorems~(\ref{entdecomp}) and (\ref{infodecomp}).}
\be
-\log Z=\sum_{i=1}^L\log p_i(s_0)-\sum_{i<j}^L\log\left[\frac{p_{ij}(s_0)}{p_i(s_0)p_j(s_0)}\right]+\cdots\;.
\ee
Inserting this into (\ref{part}) we can now identify
\be
\la h_i\ra &=&H(i)+\log p_i(s_0)\nonumber\;, \\
\la J_{ij}\ra&=&-H(i:j)-\log\left[ \frac{p_{ij}(s_0)}{p_i(s_0)p_j(s_0)}\right]\;.
\ee
This implies that in the BvH model, there is a relationship between the energy contribution to a site $\la h_i\ra$ and the entropy $H(i)$ of that site, and it is precisely the relationship that is summarized by the thermodynamical Helmholtz free-energy relation
\be
F=E-T S\;,
\ee
where $F=-\log Z$ is the free energy, $E$ is the average energy, and $S$ is the Gibbs entropy ($T$, of course, is the temperature). As a consequence, we see that the parameters of the Potts model are given entirely by quantities that can be extracted directly from the MSA, without the need of any optimization:
\be
h_i(s_a)&=&M^{(1)}_{ia}\;, \\
J_{ij}(s_a,s_b)&=&M^{(2)}_{ijab}\;.
\ee
Thus, the energy of a particular sequence ${E}(\vec s)$ can now be calculated from the first- and second-order PWMs (just as in BvH theory), without fitting any parameter to a model. 
In the next section, we extend this approach to take into account arbitrary-order correlations between monomers, in order to construct predictive energy and information scores for arbitrary sequences.

\section{Information and energy scores}
Using the PWM (\ref{pwm}) obtained from an alignment of functional sequences in an ensemble ${\cal S}_e$, we can define the first-order energy score of an {\em arbitrary} sequence $\vec s^{(k)}$ as 
\be
{E}_1(\vec s^{(k)})=\tr(M^{(1)}G_k)\;, \label{e1}
\ee 
where $G_k$ is the $D\times L$ matrix representing sequence $\vec s^{(k)}$, which has a `1' at position $i$ if symbol $a$ is found there, and a `0' otherwise. With this definition, the consensus sequence $\vec s_0$ has $E_1(\vec s_0)=0$, and sequences that carry symbols at that position that rarely occur in the MSA score badly (have high energy). For cases where the symbol does not occur at all in the MSA at that position, we have to introduce the concept of pseudocounts in order to deal with zeros in the denominator of (\ref{pwm}), as we discuss in more detail below.

The average first-order energy score over the ensemble ${\cal S}_e$ then becomes
\be
\la E_1\ra_e=H_1(S|e)+\sum_i^L\log p_0(s_i)\;,
\ee
where $H_1(S|e)$ is the first-order approximation to the Shannon entropy of the functional ensemble ${\cal S}_e$, defined in (\ref{ent1}). This is the first-order version of the Helmholtz free energy relationship.

In preparation of a discussion of information scores, let us also introduce the first-order information matrix\footnote{This is in fact Schneider's information score introduced in~\cite{Schneider1997}, but with a uniform prior $1/D$ and without finite sample size correction, which we will take care of using pseudocounts.}
\be
R^{(1)}_{ia}=\log\left[\frac{p_i(s_a)}{1/D}\right]\;,
\ee
defined in such a way that its average is the first-order Shannon information (if we take logs to the base of the alphabet $D$)
\be
\la R_1\ra_e=L-\sum_i^LH(i)\;.
\ee
The corresponding information score for sequence $\vec s_k$ is
\be
R_1(\vec s^{(k)})=\tr(R^{(1)}G_k)\;, \label{r1}
\ee
We now construct energy and information scores that include higher-order corrections. In general,
\be
E(s_1\cdots s_L)=\sum_{i=1}^L E(s_i )-\sum_{i<j}^LE(s_i:s_j)+\sum_{i<j<k}^L E(s_i:s_j:s_k)-\cdots +(-1)^{L-1}E(s_i:s_j:\cdots s_L)\;. \label{exact-e}
\ee
Here, 
\be
E(s_i)&=&\log{\frac{p_i(s_0)}{p_i(s_i)}}\;,\\
E(s_is_j)&=&\log\frac{p_{ij}(s_0)}{p_i(s_i)p_j(s_j)}\;\\
E(s_i:s_j)&=&E(s_i)+E(s_j)-E(s_is_j)\;. \label{en2}
\ee
Analogous relations hold for information scores for monomers, dimers, and so on. 
We now construct approximations to the exact energy score (\ref{exact-e}). The first-order approximation was already given in (\ref{e1}), but we rewrite it here:
\be  \label{entscore}
{E}_1(s_1\cdots s_L)&=&\sum_{i=1}^LE(s_i)\;.\\
{E}_2(s_1\cdots s_L)&=&\sum_{i=1}^LE(s_i)-\sum_{i<j}^LE(s_i:s_j)\;,\\
E_3(s_1\cdots s_L)&=&\sum_{i=1}^L E(s_i )-\sum_{i<j}^LE(s_i:s_j)+\sum_{i<j<k}^L E(s_i:s_j:s_k)
\ee
We can rewrite the correction terms as a function of joint energy scores using the decomposition theorem. For example, using (\ref{en2}), we find
\be
E_2(s_1\cdots s_L)=\sum_{i=1}^LE(s_i)-\sum_{i<j}^L(E(s_i)+E(s_j)-E(s_is_j))=\sum_{i<j}^LE(s_is_j)-(L-2)\sum_{i=1}^LE(s_i)\;.
\ee
Likewise, if we neglect fourth-order corelations,
\be
E_3(s_1\cdots s_L)=\sum_{i<j<k}^LE(s_is_js_k)-(L-3)\sum_{i<j}^LE(s_is_j)+\frac12(L-2)(L-3)\sum_{i=1}^LE(s_i)\;.
\ee
Turning now to information scores (here, $R(s_i)=\log\left[ D p_i(s_i)\right]$, $R(s_is_j)=\log\left[ D^2p_{ij}(s_is_j)\right]$ and so on, and logs are to the base $D$) we can define
\be \label{infos}
I_1(s_1\cdots s_L)&=&\sum_{i=1}^LR(s_i)\;.\\
I_2(s_1\cdots s_L)&=&\sum_{i=1}^LR(s_i)-\sum_{i<j}^LR(s_i:s_j)\;,\\
I_3(s_1\cdots s_L)&=&\sum_{i=1}^LR(s_i)-\sum_{i<j}^LR(s_i:s_j) +\sum_{i<j<k}^L R(s_i:s_j:s_k)\;,
\ee
where the shared information scores $R(s_i:s_j)$ etc.\ are defined analgously to the shared energy scores. Averaging these over the sequences in the functional ensemble ${\cal S}_e$ will return the approximations (\ref{ent1}), (\ref{ent2}), and so on.

\section{Application to sequences of self-replicating programs}
In the previous section we saw that energy and information scores are built up hierarchically, adding pairwise, three-part, and higher correlations until the full functionality emerges. But what level of correlations do we need to keep track of in order to accurately capture a sequence's functionality? To test this, we will use a data set of sequences in which correlations are important, but is also large enough so that higher-order PWMs are statistically significant. Previously, we analyzed a set of 36,171 sequences of self-replicating computer programs from the digital life system Avida (see, e.g.,~\cite{Adami1998,Adami2006,Ofriaetal2009}. Avida simulates a computer within a standard computer, within which self-replicating programs written in a custom programming language live and thrive. The programming language typically uses only 26 instructions (conveniently labeled by the lowercase letters a-z) that are chosen in such a way that programs are highly evolvable. Because the replication of instructions is noisy (giving rise to mutations), populations of programs evolve and adapt to user-specified fitness landscapes. 
The $N_e=36,171$ sequences that we analyze here represent the complete set of all possible self-replicating programs of length $L=9$ (out of a possible $26^9\approx 5.43\times 10^{12}$) obtained via an exhaustive search~\cite{CGAdami2021}. ``Self-replication" is defined here as the capacity to form a growing colony of programs (in the absence of mutations) and does not necessarily require perfect copying of information. Understanding the structure of the fitness landscape of self-replicators is an important element in trying to understand how replication could have emerged from non-replication, i.e., the origin of life~\cite{AdamiLaBar2017,Nitashetal2017}.

\begin{figure}[!h]
\centering\includegraphics[width=0.25\textwidth]{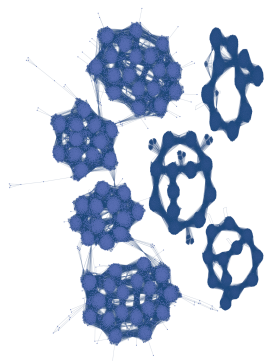}
\caption{The four largest connected components of the network of self-replicators comprise 31,592 sequences (about 87\% of the total. In these networks, two replicators are joined by an edge if they differ by exactly one point-mutation.}
\label{fig1}
\end{figure}

The 36,171 sequences form complex clusters in genotype space (see Fig.~\ref{fig1}). 
We can immediately calculate the information content of these 9-mers using Szostak's formula (\ref{fun}),
which gives\footnote{Choosing a base for the logarithm gives units to entropy and information. In the following, we will take logs to the base of the alphabet, giving units of ``mers". An $n$-mer has $n$ mers of potential information (entropy).}
\be
I_9=-\log_{26}{\frac{36,171}{26^9}}\approx 5.77\  {\rm mers}\;\label{exact}
\ee
that is, about 64\% of the sequence is information (the smallest possible replicator in this world has $L=8$, with an even more compressed code~\cite{CGAdami2021}.

\subsection{Emergence of information from multi-partite correlations}
\begin{figure}[!h]
\centering\includegraphics[width=0.4\textwidth]{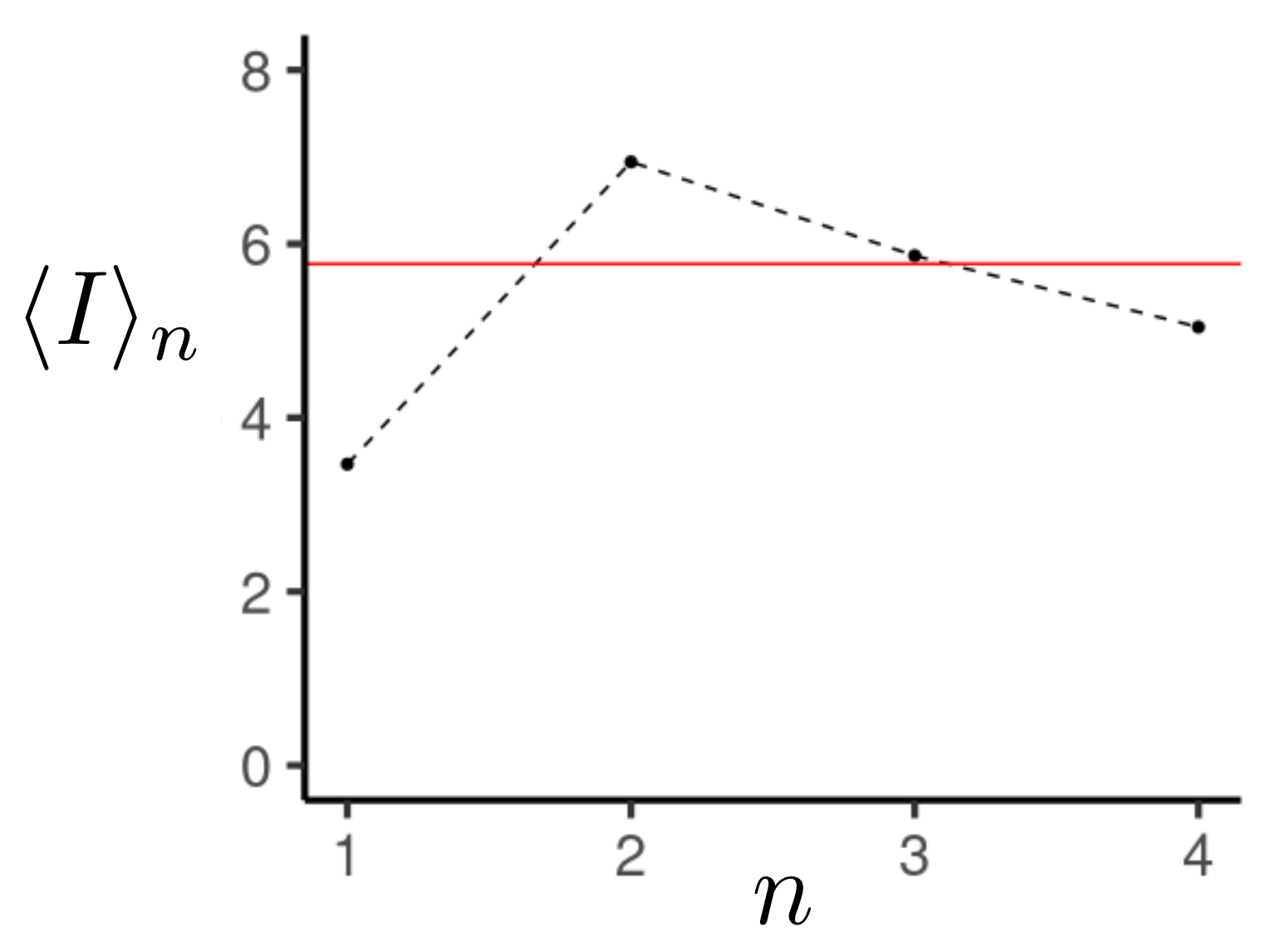}
\caption{Mean information scores (Shannon information) as a function of the degree of correlations taken into account. $\langle I_1\rangle$ corresponds to the information disregarding pair-wise correlations between sites, while $\langle I\rangle_2$ disregards third-order correlations, and so on. The red line is the exact information $I_9=5.77$ mers.}
\label{fig2}
\end{figure}

We can now ask how this information emerges from correlations, by averaging the information scores (\ref{infos}) over $S_e$. In Fig.~\ref{fig2} we show the approximations to the exact information score (\ref{exact}) as a function of the order $n$ of the approximation.
As expected from the information decomposition theorem, the approximations move closer to the exact value as more correlations are incorporated.

\subsection{Prediction of function from information scores}
In this section we will test how much ``correlation information" (information stored in correlations between sites, as opposed to bias in per-site frequency) is necessary to distinguish functional from non-functional sequences. Because we will now be scoring sequences that are {\em not} in the set ${\cal S}_e$, we have to introduce pseudocounts in the PWMs. As is clear from (\ref{pwm}), if a particular symbol $s_a$ at position $i$ does not occur in the MSA, then the maximum likelihood estimator of the probability $p_i(s_a)$ vanishes, leading to an infinite matrix element. As a consequence, the score of a sequence that contains such a symbol at the position is not well-defined. A common cure for this defect is to introduce pseudocounts, which add a constant to every possible symbol count. While the common Laplace pseudocount adds 1 to each count ($n_a\to n_a+1$ and therefore $p_i(s_a)\to (n_a+1)/(N_e+D)$,
the pseudocount is often used as a variable that can be adapted to the statistics of the dataset in question (see, e.g.,~\cite{Nemenmanetal2002,Nemenman2011}). In what follows, we adopt a pseudocount of $10^{-6}$ throughout, but note that the results on classification accuracy scale only extremely weakly over the 12 orders of magnitude of pseudocounts we tested. Generally speaking, an infinitesimal pseudocount moves information scores of un-represented symbols far out into the negative, and energy scores far to the positive. The pseudocount for higher-order PWMs (we construct those explicitly up to $n=4$, where $n$ is the order of correlations that are included) is adjusted so that the marginal probabilities remain correct.

In order to predict the function of an arbitrary sequence in light of an MSA of $N\leq N_e$ functional ones, we need to construct a classifier in such a manner that, given a particular threshold $\tau$, the score divides functional from non-functional sequences. For energy scores, we would deem those with a score $E(\vec s)<\tau$ as functional, while for information scores we would deeem those with $I(\vec s)>\tau$ to be replicators. Note that because we have exhaustively classified this dataset already, we can determine the accuracy of classification (given a subset of sequences $N<N_e$ for a given classifier) with perfect accuracy.

In principle, we can construct classifiers directly from the information score functions (\ref{infos}), but due to the fact that the correction terms to $I_1$ involve the summation of many terms with alternating signs, the pseudodocount prevents these classifiers from being effective\footnote{This holds equally when using energy score functions~(\ref{entscore}).}. Indeed, it appears to be impossible to construct pseudocounts in a manner that does not introduce spurious correlations at all levels $n$ (while it is of course possible to make sure no spurious correlations are introduced for a chosen given level). However, it turns out that sums over the multiple-site energy and information functionals $E(s_is_j\cdots)$ and $R(s_is_j\cdots)$ do work well, because they do not rely on cancellations. For example, we can define the following information classifiers (similar classifiers can be built from energy functionals):
\be
D_1(\vec s)&=&\sum_{i=1}^L R(s_i)  \label{d1}\;,\\
D_2(\vec s)&=&\sum_{i<j}^LR(s_is_j)  \label{d2}  \;,\\  
D_3(\vec s)&=&\sum_{i<j<k}^LR(s_is_js_k)  \label{d3} \;,\\
D_4(\vec s)&=&\sum_{i<j<k<m}^LR(s_is_js_ks_m)   \label{d4}  \;.
\ee
To test the quality of the classifiers, we create test data containing $10^5$ sequences. Specifically, we create three ensembles: ${\cal S}_R$ with randomly generated sequences (all sequences are $L=9$) that we made sure are all non-functional, as well as a set of one-mutants ${\cal S}_1$ and two-mutants ${\cal S}_2$ that are one or two mutations away from functional sequences, but known not to be functional. We expect the one-mutant set in particular to be the hardest to classify.  

\begin{figure}[!h]
\centering\includegraphics[width=0.8\textwidth]{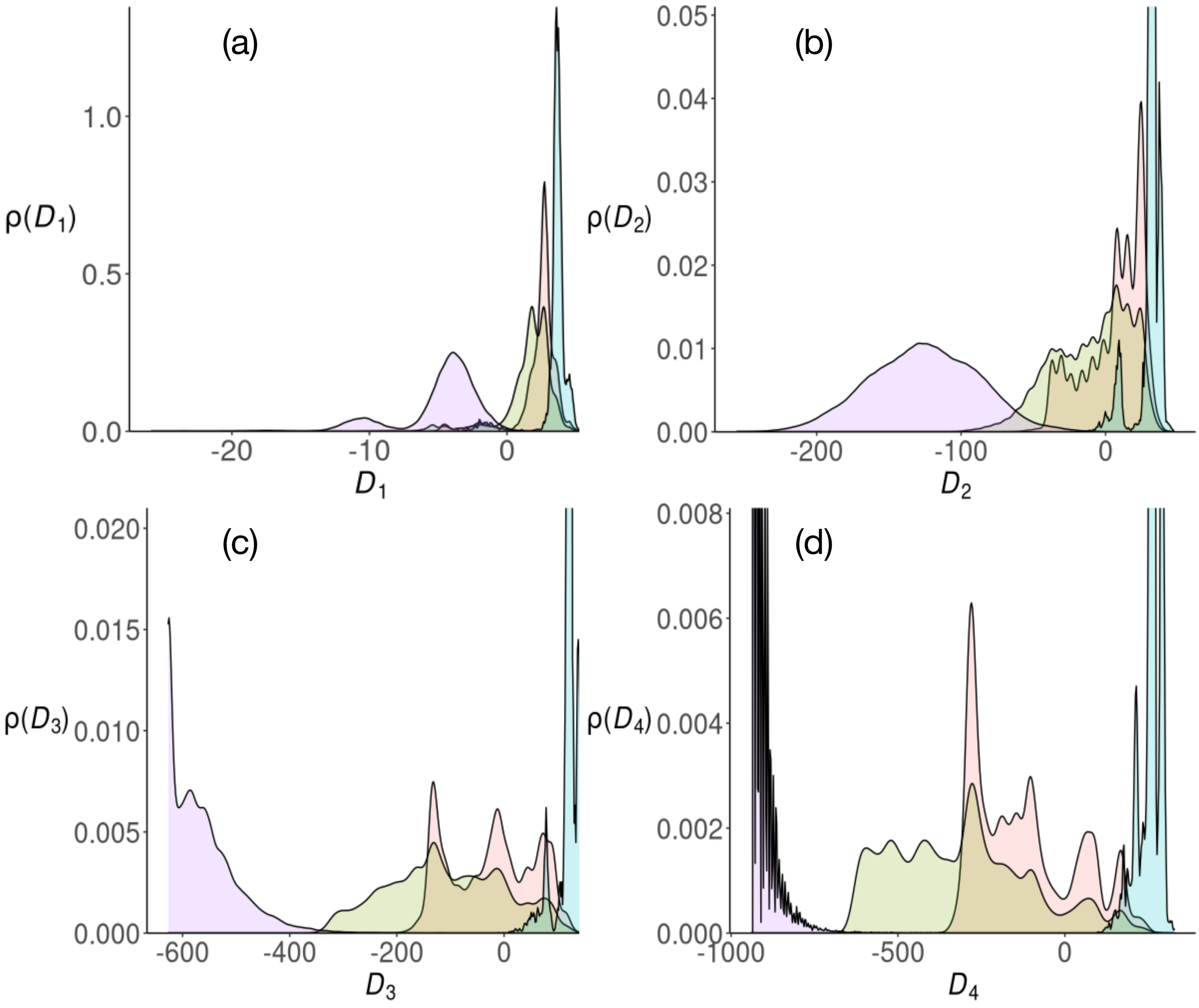}
\caption{Density plots of scores generated by the classifiers (\ref{d1}) to (\ref{d4}), with PWMs using all sequences in ${\cal S}_e$ and a pseudocount $10^{-6}$. (a) Density of scores obtained with $D_1$ for true sequences (${\cal S}_e$, light blue), false sequences (${\cal S}_R)$, purple), single mutant false (${\cal S}_1$, red), and double-mutant false (${\cal S}_2$, green). Note that for visibility, some distributions are cropped at the top..
(b) Density of scores for classifier $D_2$. (c) Density of scores for classifier $D_2$. Density of scores for classifier $D_3$.(d) Density of scores for classifier $D_4$.}
\label{fig3}
\end{figure}

We show typical density distributions of scores for the classifiers in Fig.~\ref{fig3}. These classifiers were all ``trained" on the full set ${\cal S}_e$, meaning, the PWMs $M^{(n)}$ were created with the entire set of replicators in the MSA. As expected, the set of random sequences has a density distribution (purple in Figs.~\ref{fig3}(a-d)) far removed from the functional sequences (light blue), but the one-mutant and two-mutant sequences have a significant overlap with functional sequences if $D_1$ is used as a classifier. Including higher-order correlations reduces this overlap, until the overlap (mis-classification) almost completely disappears in $D_4$. 

\begin{figure}[!h]
\centering\includegraphics[width=0.7\textwidth]{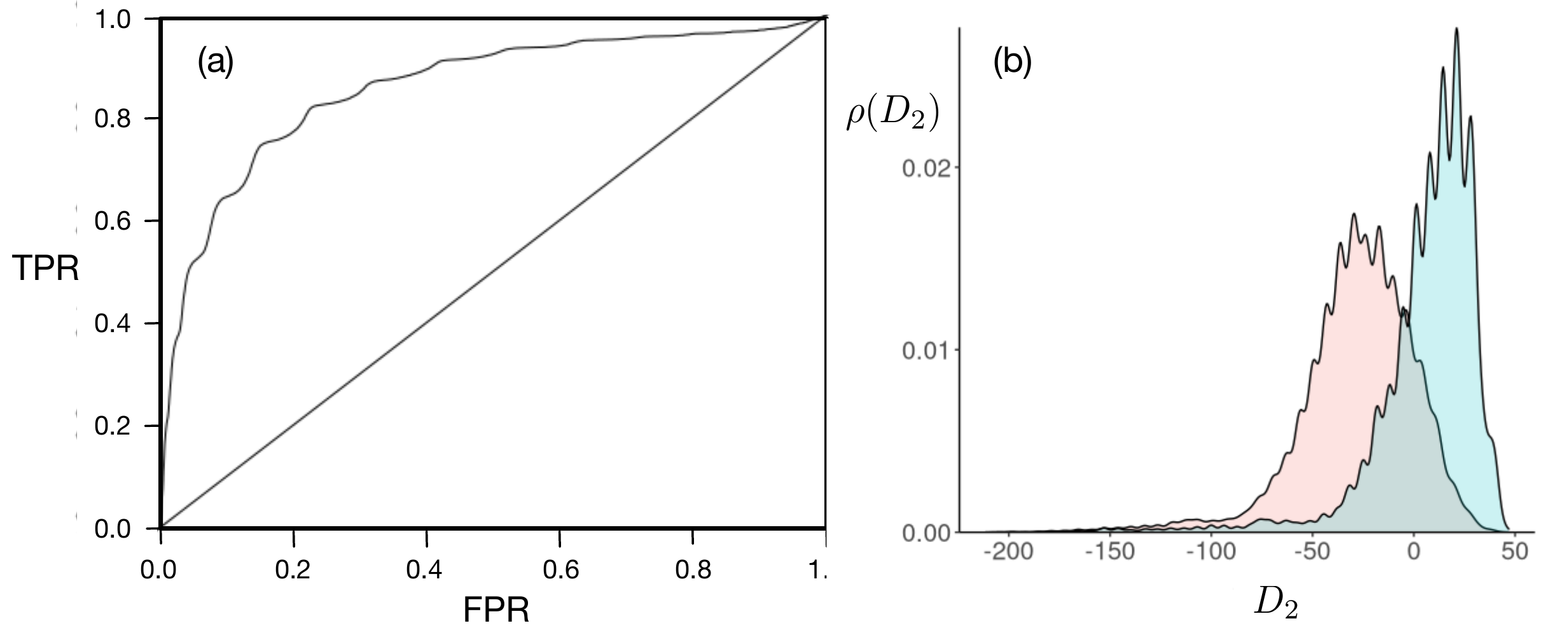}
\caption{(a) ROC plot for classifier $D_2$, constructed from 1\% of the $N_e$ sequences in ${\cal S}_e$, and evaluated on 100,000 uniformly sampled non-functional single-point mutants of functional sequences, as the threshold parameter $\tau$ is varied from $\tau=45$ to $\tau=-211$. True positives are functional. The diagonal line is the ROC of a random (unspecific) classifier. (b) The distribution of functional sequences ${\cal S}_e$ (blue) and of the non-functional single-mutant sequences ${\cal S}_1$, both evaluated by $D_2$.
}
\label{fig4}
\end{figure}

We can characterize the accuracy of the classifier as a function of the size of the training set by measuring the quality score of an ROC (receiver operating chararacteristic, see, e.g.,~\cite{Fawcett2006}) plot. ROCs are generated by plotting the false positive rate (FPR) against the true positive rate (TPR) as a function of the threshold $\tau$ in the classifier. A typical ROC (for classifier $D_2$ trained on 1\% of ${\cal S}_e$ used on 100,000 non-functional single-point mutants of the functional set) is shown in Fig.~\ref{fig4}. The quality score $Q$ of the classifier is obtained by calculating the area under the curve (AUC) of the classifier, as compared to the AUC=0.5 of a random classifier where the TPR increases linearly with the FPR.

\begin{figure}[!h]
\centering\includegraphics[width=\textwidth]{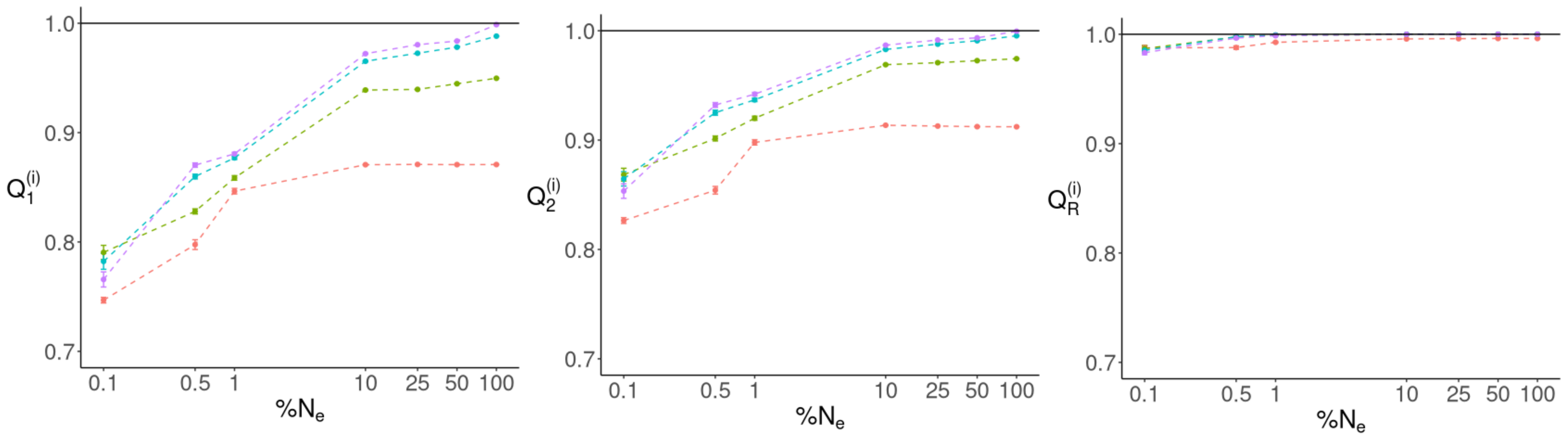}
\caption{Quality of classifiers (AUC of the ROC curves) when testing against different test sets, as a function of the traning set size (in percentage of the full set of size $N_e$). Note log scale. In all panels, red points and lines refer to the first-order classifiers $D_1$, green dots and lines depict the performance of second-order classifiers $D_2$, while the third and fourth-order classifiers are in blue, and purple, respectively. Error bars are standard error over ten replicates (except for $N=N_e$).
(a) $Q_1$ is the quality score of classifiers for a test set made from 31,291 non-replicating one-mutants of the functional set ${\cal S}_2.$ (b) $Q_2$ quantifies the quality of the classifiers when tested on the two-mutant non-functional set ${\cal S}_2$. (c) $Q_R$ measures the performance of the classifiers when tested on 31,291 randomly generated non-functional of 9-mers.  
}
\label{fig5}
\end{figure}
It is clear from Fig.~\ref{fig5} that all classifiers work well even when only 0.1\% of the true replicators (36 sequences) are used to create the MSA, but for training sets that small, the classifiers $D_3$ and $D_4$ that use three-partite and four-partite correlations perform less well than $D_1$ and $D_2$ on the test data for one- and two mutants of functional sequences, which can be attributed to overfitting. Indeed, for data sets this small, the equilibrium assumption behind the max entropy approach will be violated for those PWMs. However, for larger MSAs, including higher-order correlations significantly improves the performance of the classifier. When using the full set of replicators, taking into account up to four-way correlations appears to capture almost all of the information encoded in the sequences. On random sequences, all classifiers perform close to perfectly even on the smallest training set size, and including multivariate correlations is clearly overkill. The quality of classifiers based on energy functionals is the same as for those described here (data not shown), as they are intimately connected via the Helmholtz free energy relation.

\section{Discussion}
Fano's entropy and information decomposition theorems allow us to understand how functional information emerges from the combination of multivariate correlations. Here we have shown that defining energy and information functionals built from multi-variate position weight matrices allow us to capture the correlations inherent in multiple sequence alignments in a model-free manner. Standard approaches to fitting MSAs rely on expensive inverse models to fit the bi-partite correlations in the data, but must reduce the alphabet size considerably to reduce the number of parameters in the Potts or Ising models. It is clear from this analysis that such a fit is entirely unnecessary, as the marginal probabilities can be extracted from the data as long as the maximum entropy assumption is used on single sites, pairs of sites, triples, and so on.  

We tested this approach on the largest computational genotype-phenotype map created to date. Having a complete genotype-phenotype map allows us to calculate the information content of functional sequences exactly, and study how this information is built up by lower-order correlations. Further, a complete data set helps in creating ensembles that can be used to test the classification accuracy of function prediction. In particular, we used the model-free
approach to construct classifiers that can distinguish functional from non-functional sequences even when the non-functional sequences are single-site polymorphisms of the functional ones. Of course, this approach needs to be tested on biological data sets with significantly longer sequences. However, since the computational cost of creating these complex classifiers scales mainly with the level of multivariate correlations that are included, we expect to be able to classify sequences of several hundreds of sites with full alphabet size (for example, $D=20$ or 21 for proteins) taking into account all tri-partite correlations. For binary data such as neuronal spike trains, we expect to be able to handle sequences of thousands of sites, possibly including even more than 4-part correlations. 

Sequence models of the Ising or Potts type perform markedly better than the standard Hidden Markov Models (HMMs) that have been developed and used since the early 1970s~\cite{Rabiner1989}, because they can take into account arbitrary correlations between pairs of symbols, rather than just those between adjacent symbols in HMMs. However, going beyond pair-wise correlations is deemed impossible for Ising models because of the associated explosion in the number of parameters. While more general pattern classification task (of which sequence classification is but a subset) can nowadays be achieved by training deep neural networks on annotated data, these methods have also displayed severe drawbacks~\cite{Nguyenetal2015,JoBengio2018}, for example by being easily fooled by detracting patterns. Furthermore, training such models on a large corpus of data is expensive. The model-free classifiers that we have described here are different from deep networks in important ways, as they do not require any training, and their vulnerability to overfitting is easily controlled by adjusting the level of correlation $n$ to be included to the size of the MSA.

\enlargethispage{20pt}

%
%
%

\noindent {\bf Acknowledgements} This work was supported in part by funds provided by Michigan State University in support of the BEACON Center for the Study of Evolution in Action. We acknowledge computational resources provided by the Institute for Cyber-Enabled Research (iCER) at Michigan State University.


\begin{thebibliography}{10}
\expandafter\ifx\csname urlstyle\endcsname\relax
  \providecommand{\doi}[1]{(doi:\discretionary{}{}{}#1)}\else
  \providecommand{\doi}{(doi:\discretionary{}{}{}\begingroup
  \urlstyle{rm}\Url)}\fi

\bibitem{Szostak2003}
Szostak JW. 2003 Functional information: Molecular messages.
\newblock \emph{Nature} \textbf{423}, 689.

\bibitem{Hazenetal2007}
Hazen RM, Griffin PL, Carothers JM, Szostak JW. 2007 Functional information and
  the emergence of biocomplexity.
\newblock \emph{Proc. Natl. Acad. Sci. USA} \textbf{104}, 8574--8581.

\bibitem{Carothersetal2004}
Carothers JM, Oestreich SC, Davis JH, Szostak JW. 2004 Informational complexity
  and functional activity of {RNA} structures.
\newblock \emph{J. American Chem. Society} \textbf{126}, 5130--5137.

\bibitem{GuptaAdami2016}
Gupta A, Adami C. 2016 Strong selection significantly increases epistatic
  interactions in the long-term evolution of a protein.
\newblock \emph{PLoS Genet} \textbf{12}, e1005960.

\bibitem{AdamiLaBar2017}
Adami C, LaBar T. 2017 From entropy to information: {Biased} typewriters and
  the origin of life.
\newblock In \emph{Information and Causality: From Matter to Life} (ed.
  S~Walker, P~Davies, G~Ellis), pp. 95--112. Cambridge, MA: Cambridge
  University Press.

\bibitem{AdamiCerf2000}
Adami C, Cerf NJ. 2000 Physical complexity of symbolic sequences.
\newblock \emph{Physica D} \textbf{137}, 62--69.

\bibitem{Adami2004}
Adami C. 2004 Information theory in molecular biology.
\newblock \emph{Phys. Life Reviews} \textbf{1}, 3--22.

\bibitem{Adami2012}
Adami C. 2012 The use of information theory in evolutionary biology.
\newblock \emph{Ann. N.Y. Acad. Sci.} \textbf{1256}, 49--65.

\bibitem{Fano1961}
Fano RM. 1961 \emph{Transmission of Information: A Statistical Theory of
  Communication}.
\newblock New York and London: MIT Press and John Wiley.

\bibitem{Sylvester1883}
Sylvester J. 1883 Note sur le th{\'e}or{\`e}me de {Legendre}.
\newblock \emph{Comptes Rendus Acad. Sci. Paris} \textbf{96}, 463--465.

\bibitem{Hinkleyetal2011}
Hinkley T, Martins J, Chappey C, Haddad M, Stawiski E, Whitcomb JM, Petropoulos
  CJ, Bonhoeffer S. 2011 A systems analysis of mutational effects in {HIV-1}
  protease and reverse transcriptase.
\newblock \emph{Nat Genet} \textbf{43}, 487--9.

\bibitem{Kouyosetal2011}
Kouyos RD, von Wyl V, Hinkley T, Petropoulos CJ, Haddad M, Whitcomb JM,
  B{\"o}ni J, Yerly S, Cellerai C, Klimkait T, G{\"u}nthard HF, Bonhoeffer S,
  {Swiss HIV Cohort Study}. 2011 Assessing predicted {HIV-1} replicative
  capacity in a clinical setting.
\newblock \emph{PLoS Pathog} \textbf{7}, e1002321.

\bibitem{Schneidmanetal2006}
Schneidman E, Berry MJ 2nd, Segev R, Bialek W. 2006 Weak pairwise correlations
  imply strongly correlated network states in a neural population.
\newblock \emph{Nature} \textbf{440}, 1007--12.

\bibitem{MoraBialek2011}
Mora T, Bialek W. 2011 Are biological systems poised at criticality?
\newblock \emph{J. Stat. Phys.} \textbf{144}, 268--302.

\bibitem{Fergusonetal2013}
Ferguson AL, Mann JK, Omarjee S, Ndung'u T, Walker BD, Chakraborty AK. 2013
  Translating {HIV} sequences into quantitative fitness landscapes predicts
  viral vulnerabilities for rational immunogen design.
\newblock \emph{Immunity} \textbf{38}, 606--17.

\bibitem{Biswasetal2021}
Biswas A, Haldane A, Levy RM. 2021.
\newblock The role of epistasis in determining the fitness landscape of {HIV}
  proteins.
\newblock BioRxiv 448646.

\bibitem{BergvonHippel1987}
Berg OG, von Hippel PH. 1987 Selection of {DNA} binding sites by regulatory
  proteins. {Statistical-mechanical} theory and application to operators and
  promoters.
\newblock \emph{J. Mol. Biol.} \textbf{193}, 723--750.

\bibitem{BergvonHippel1988}
Berg OG, von Hippel PH. 1988 Selection of {DNA} binding sites by regulatory
  proteins {II. The} binding specificty of cyclic {AMP} receptor protein to
  recognition sites.
\newblock \emph{J. Mol. Biol.} \textbf{200}, 709--723.

\bibitem{Stormo2000}
Stormo GD. 2000 {DNA} binding sites: Representation and discovery.
\newblock \emph{Bioinformatics} \textbf{14}, 16--23.

\bibitem{BrownCallan2004}
Brown CT, {Callan Jr} CG. 2004 {Evolutionary comparisons suggest many novel
  {cAMP} response protein binding sites in {\em {E}scherichia coli}}.
\newblock \emph{Proc. Natl. Acad. Sci. USA} \textbf{101}, 2404--2409.

\bibitem{CliffordAdami2015}
Clifford J, Adami C. 2015 Discovery and information-theoretic characterization
  of transcription factor binding sites that act cooperatively.
\newblock \emph{Phys Biol} \textbf{12}, 056004.

\bibitem{Schneider1997}
Schneider TD. 1997 Information content of individual genetic sequences.
\newblock \emph{J. theor. Biol.} \textbf{189}, 427--441.

\bibitem{Adami1998}
Adami C. 1998 \emph{Introduction to Artificial Life}.
\newblock New York: Springer Verlag.

\bibitem{Adami2006}
Adami C. 2006 Digital genetics: Unravelling the genetic basis of evolution.
\newblock \emph{Nat Rev Genet} \textbf{7}, 109--118.

\bibitem{Ofriaetal2009}
Ofria C, Bryson DM, Wilke CO. 2009 Avida: A software platform for research in
  computational evolutionary biology.
\newblock In \emph{Artificial Life Models in Software} (ed. M~Komosinski,
  A~Adamatzky), pp. 3--35. Springer London.

\bibitem{CGAdami2021}
{C G} N, Adami C. 2021 Information-theoretic characterization of the complete
  genotype-phenotype map of a complex pre-biotic world.
\newblock \emph{Phys Life Rev} \textbf{38}, 111--114.

\bibitem{Nitashetal2017}
{C G} N, LaBar T, Hintze A, Adami C. 2017 Origin of life in a digital
  microcosm.
\newblock \emph{Philos Trans R Soc Lond A} \textbf{375}, 20160350.

\bibitem{Nemenmanetal2002}
Nemenman I, Shafee F, Bialek W. 2002 Entropy and inference, revisited.
\newblock In \emph{Adv Neural Inf Process Syst. vol. 14} (ed. G~Dietterich,
  S~Becker, Z~Ghahramani), pp. 471--478.

\bibitem{Nemenman2011}
Nemenman I. 2011 Coincidences and estimation of entropies of random variables
  with large cardinalities.
\newblock \emph{Entropy} \textbf{13}, 2013--2023.

\bibitem{Fawcett2006}
Fawcett T. 2006 An introduction to {ROC} analysis.
\newblock \emph{Pattern Recognition Letters} \textbf{27}, 861--874.

\bibitem{Rabiner1989}
Rabiner LR. 1989 A tutorial on {Hidden Markov} models and selcted applications
  in speech recognition.
\newblock \emph{Proceeding of the IEEE} \textbf{77}, 257--286.

\bibitem{Nguyenetal2015}
Nguyen A, Yosinski J, Clune J. 2015 Deep neural networks are easily fooled:
  High confidence predictions for unrecognizable images.
\newblock In \emph{Computer Vision and Pattern Recognition (CVPR '15)}. IEEE.

\bibitem{JoBengio2018}
Jo J, Bengio Y. 2018 Measuring the tendency of {CNNs} to learn surface
  stastistical regularities.
\newblock ArXiv:1711.11561

\end{thebibliography}

\end{document}